# On Almost Controllability of Dynamical Complex Networks with Noises

CAI, Ning[1, 2, 3]  HE, Ming[4]  WU, Qiu-Xuan[2]  KHAN, M. Junaid[5]

[1] College of Electrical Engineering, Northwest Minzu University, Lanzhou, China
[2] School of Automation, Hangzhou Dianzi University, Hangzhou, China
[3] Key Laboratory of China's National Linguistic Information Technology, Lanzhou, China
[4] High-Tech Institute of Xi'an, Xi'an, China
[5] PN Engineering College, National University of Sciences and Technology, Islamabad, Pakistan

**Abstract:** This paper discusses the controllability problem of complex networks. It is shown that almost any weighted complex network with noise on the strength of communication links is controllable in the sense of Kalman controllability. The concept of almost controllability is elaborated by both theoretical discussions and experimental verifications.

**Key Words:** Complex network; Controllability; Weight; Communication link

## 1. Introduction

In a dynamical network, the vertices may contain certain time-varying information or substance. Such a network is controllable when expected quantity of information or substance on any vertices can be achieved, if only with appropriate external input. Observability and controllability are dual alternatives. Integrated with stability, they form the theoretical foundation for most of the systems analysis and synthesis problems. Thus, study of controllability has become one of the most important subjects in systems science.

Since the start of this century, the controllability problems of dynamical distributed large-scale systems have intrigued many scholars from both the control [1-9] and the theoretical physics [10-17] communities, and are expected to keep on attracting the attention of more and more disciplines.

Tanner [1] addressed the controllability of systems with a single leader and conjectured that excessive connectivity may be detrimental to controllability, by giving a definition of graph controllability based on partitioning of the associated Laplacian matrix. Considering the correlation between the levels of graph symmetry and controllability, Rahmani and Mesbahi [2] further extended the results in [1]. Cai *et al.* [3-5] concerned the controllability problems of a class of high-order systems, presenting a scheme for improving controllability based on certain equivalent transformation of the network topologies. Liu *et al.* [6] studied the controllability of discrete-time systems with switching graph topologies. Ji *et al.* [7-8] addressed the interactive protocols, in an endeavor to integrate the influence of three factors upon controllability, i.e. the protocol, the vertex dynamics and the network topology. Guan *et al.* presented topological criteria to check the controllability of directed networks via introducing a concept of leader-follower connectedness [9].

[1] Corresponding author: Cai, Ning (caining91@tsinghua.org.cn)

Liu *et al.* [10] dealt with the structural controllability of complex networks. They selected an index to quantitatively measure the level of controllability of complex networks, which is based on the least amount of requisite independent input signals. Along the route of [10], there have emerged a number of papers, mainly from the theoretical physics community, e.g. [11-17]. Particularly, Yan *et al.* [11] concentrated upon the problem of minimal energy cost for maneuvering the vertices. Yuan *et al.* [12] concerned the exact controllability of undirected networks with identical edge weights and discovered certain consistency between structural controllability and exact controllability. Sun and Motter [13] discovered a fact that even the systems that pass the Kalman controllability criteria may still be uncontrollable in practice due to numerical effects, which was later substantiated in [14].

The concept of controllability for dynamical systems was initially proposed by Kalman, along with a set of algebraic criteria to check whether any given system is controllable. Kalman perspective of controllability has formed the foundation of the controllability/observability theory in systems science. Actually, there exists one essential problem for this concept, i.e. almost all real-world dynamical systems are completely controllable in the sense of Kalman controllability. Such a problem becomes especially prominent in dynamical complex networks as compared with conventional dynamical systems, because 1) the order of a dynamical complex network is usually ultra-high; 2) modeling uncertainties are very common for networked systems. Nonetheless, this problem is often neglected, especially by scholars from the theoretical physics community, e.g. [12]. Additionally, the knowledge that almost controllability is an essential nature of complex networks may act as a theoretical foundation for extending the study of controllability, from qualitative to quantitative [18]. Thus, it would be enlightening to expound and emphasize it specifically.

The primary objective of this paper is an endeavor to elucidate the concept of almost controllability and facilitate a deeper understanding of the controlling mechanism of dynamical networks.

The main contribution is elaborating that unless there is precise ideal modeling, for real-world complex networks almost every network is controllable in the sense of Kalman controllability, and besides, any random infinitesimal perturbation could lead to a controllable network. Discussions about almost controllability will be launched on two concrete algebraic representations of complex networks respectively: adjacency matrix and Laplacian matrix. The analysis will be conducted both theoretically and empirically, or in another word, in mathematical and physical manners.

The rest of this paper is organized as follows. Section 2 formulates the models and the controllability problem. Section 3 expounds the almost controllability theoretically, from various aspects. Section 4 verifies the theoretical analysis by numerical experiments. Finally, Section 5 presents a brief conclusion.

## 2. Problem Formulation and Preliminaries

Two types of dynamical networks are concerned in this paper.

### 2.1. Dynamical network represented by adjacency matrix

The dynamic motion of the first type network systems is described as:

$$\dot{x}_i = \frac{dx_i(t)}{dt} = \sum_{j=1}^{N} a_{ij} x_j + u_i \quad (i = 1, 2, \ldots, N) \tag{1}$$

where $x_i \in R$ denotes the time-varying state of dynamical vertex *i*; $u_i \in R$ denotes the input signal on vertex *i*; and $a_{ij} \in R$ denotes the weight of directed arc from vertex *j* to *i*.

The system of equations (1) can be combined into a compact vectored form as:

$$\dot{x} = Ax + u \tag{2}$$

where $x(t) = [x_1(t) \quad x_2(t) \quad \cdots \quad x_N(t)]^T \in R^N$ represents the state vector of the dynamical vertices, $u(t) = [u_1(t) \quad u_2(t) \quad \cdots \quad u_N(t)]^T \in R^N$ represents the input vector, and $A = [a_{ij}] \in R^{N \times N}$ represents the weighted adjacency matrix of the network topology.

In this network, the vertices are classified into two categories: leaders and followers. Without loss of generality, suppose that the first $N_f$ vertices are followers and the remaining $N_l$ vertices are leaders, with $N_f + N_l = N$. The dynamics of leaders are driven by external signals, whereas the followers receive no external stimulus, i.e. $u_i(t) \equiv 0$ ($i$ = 1, 2, …, $N_f$). (2) can be decomposed below according to the leader-follower configuration:

$$\begin{bmatrix} \dot{x}_f \\ \dot{x}_l \end{bmatrix} = \begin{bmatrix} A_{ff} & A_{fl} \\ A_{lf} & A_{ll} \end{bmatrix} \begin{bmatrix} x_f \\ x_l \end{bmatrix} + \begin{bmatrix} 0 \\ u_l \end{bmatrix} \tag{3}$$

In the above equation, $A_{ff}$ denotes the subgraph which is merely composed of followers; $A_{ll}$ denotes the subgraph composed of leaders; and $A_{lf}$ & $A_{fl}$ express the directed arcs between the leaders and followers.

The network is Kalman controllable if the values of the states can be completely controlled by external inputs, i.e. the network is able to achieve any expected states within a finite time span if only with appropriate input signals, otherwise it is said to be uncontrollable. Below is the conventional definition of Kalman controllability.

*Definition 1:* A dynamical network (2) is completely controllable if for any initial values of vertex states $x_1(0), x_2(0), ..., x_N(0) \in R$, there exist $\tau < \infty$ and proper input signals $u_1(t), u_2(t), ..., u_N(t)$ $(t \in [0, \tau])$ such that $x_1(\tau) = x_2(\tau) = ... = x_N(\tau) = 0$.

Lemma 1 below provides the most fundamental criterion for checking Kalman controllability, known as the rank test.

*Lemma 1 [19]:* A dynamical system $\dot{\xi} = M\xi + B\upsilon$ with $\xi \in R^n$ being the state vector and $\upsilon \in R^m$ the input vector is completely controllable if and only if the controllability matrix is of full rank, which is

$$Co = \begin{bmatrix} B & MB & M^2B & \cdots & M^nB \end{bmatrix}$$

*Remark 1:* The controllability of system $\dot{\xi} = M\xi + B\upsilon$ is equivalent to that of the matrix pencil ($M$, $B$) because the controllability is exclusively determined by the pair of matrices $M$ and $B$.

According to (3), the overall network is decomposed into two subsystems, which are the follower subsystem $\dot{x}_f = A_{ff}x_f + A_{fl}x_l$ and the leader subsystem $\dot{x}_l = A_{ll}x_l + A_{lf}x_f + u_l$.

*Proposition 1:* The leader subsystem is always controllable.

*Proof:* Let $u_l = -A_{lf} x_f + v$, then the dynamic equation of leader subsystem can be rewritten as

$$\dot{x}_l = A_{ll} x_l + v = A_{ll} x_l + I_{N_l} v$$

The controllability matrix is

$$Co = \begin{bmatrix} I_{N_l} & A_{ll} I_{N_l} & A_{ll}^2 I_{N_l} & \cdots & A_{ll}^{N_l} I_{N_l} \end{bmatrix}$$

Observably, $rank(Co) = N_l$ and the subsystem is deemed controllable according to Lemma 1. □

Since the leader subsystem is always controllable, system (2) is Kalman controllable if and only if the follower subsystem is controllable.

*Definition 2:* A dynamical complex network (2) is said to be controllable if the matrix pencil $(A_{ff}, A_{fl})$ is controllable.

## 2.2. Dynamical network represented by Laplacian matrix

The dynamics of the second type of dynamical complex network is described as:

$$\dot{x}_i = \frac{dx_i(t)}{dt} = \sum_{j=1}^{N} a_{ij}(x_j - x_i) + u_i \quad (i = 1, 2, \ldots, N) \tag{4}$$

where $x_i \in R$ denotes the time-varying state of dynamical vertex $i$; $u_i \in R$ denotes the input signal on vertex $i$; and matrix $A = [a_{ij}]$ denotes the weighted adjacency matrix of the graph topology.

The system of equations (4) can also be integrated into vector form similar to (2), which is:

$$\dot{x} = -Lx + u \tag{5}$$

where $L$ is the Laplacian matrix of the graph topology.

In comparison with (2), the only difference of (5) is replacing the matrix '$A$' by '$-L$', with the counterpart of (3) being

$$\begin{bmatrix} \dot{x}_f \\ \dot{x}_l \end{bmatrix} = \begin{bmatrix} L_{ff} & L_{fl} \\ L_{lf} & L_{ll} \end{bmatrix} \begin{bmatrix} x_f \\ x_l \end{bmatrix} + \begin{bmatrix} 0 \\ u_l \end{bmatrix}$$

Accordingly, the following definition naturally arises.

*Definition 3:* A dynamical complex network (5) is said to be controllable if the matrix pencil $(L_{ff}, L_{fl})$ is controllable.

*Remark 2:* The state equation (2) is actually identical to a typical LTI (Linear Time-Invariant) state space model, only with an ultra-high order, and the overall dynamics is regulated by the adjacency matrix *A*. Such a model of dynamical complex networks is simpler, usually being addressed by scholars from the fields of theoretical physics, e.g. [10, 12]. Differently, the reason why state equation (5) is based on Laplacian matrix is that it is derived from the interactive protocol in (4) being built upon the relative states between the nearest neighboring vertices, in contrast with the absolute states of the neighbors in (1). The model (5) is often addressed by scholars from the control theory community, especially when dealing with the well-known consensus problems [20].

## 2.3. Relevant preliminaries

This subsection provides some important foundation underlying the theoretical discussions in the subsequent section.

*Lemma 2:* The maximum algebraic multiplicity of eigenvalues for any real-valued square matrix $M \in R^{n\times n}$ with linear constraints is almost 1, i.e. the Lebesgue measure of the set of matrices in $R^{n\times n}$ with algebraic multiplicity being higher than 1 is zero, even with linear constraints on the element values.

For self-containment of the current paper, a detailed proof of the above lemma is attached as the appendix. The analysis is analogous to the decouplability problem for complex networks [21].

*Remark 3:* The multiplicities of eigenvalues are extremely dependent on the precision of the element values of a matrix.

*Lemma 3:* (Degree of Minimal Polynomial) Suppose that matrix $M \in R^{n\times n}$ has $\mu$ distinct eigenvalues $\lambda_1, \lambda_2, ..., \lambda_\mu$, each with algebraic multiplicity $\sigma_i$ and geometric multiplicity $\zeta_i$ respectively. The Jordan canonical form of *M* is

$$\begin{bmatrix} J_1 & & & \\ & J_2 & & \\ & & \ddots & \\ & & & J_\mu \end{bmatrix} \tag{6}$$

where

$$J_i = \begin{bmatrix} J_{i1} & & \\ & \ddots & \\ & & J_{i\zeta_i} \end{bmatrix} \quad (i = 1, 2, ..., \mu) \tag{7}$$

with each Jordan block $J_{ik} \in C^{d_{ik} \times d_{ik}}$ $(k = 1, 2, ..., \zeta_i$ and $\sum_{k=1}^{\zeta_i} d_{ik} = \sigma_i)$. The degree of the minimal polynomial $\phi_M(\lambda)$ of $M$ is $\sum_{i=1}^{\mu} \max_{1 \le k \le \zeta_i} \{d_{ik}\}$.

## 3. Theoretical Analysis

The continuous-valued weight of an arc in a network reflects the strength of communication link. For real-world networks, deviations always exist for the values of weights because there exists no perfectly accurate model.

Actually, almost every weighted dynamical complex network depicted by (1) or (4) is controllable in the sense of Kalman controllability, if only there is perturbation on the arc weights. The connotation of almost controllability contains multiple layers of meanings.

Let us first consider a scenario. Suppose an existing complex network comprises only follower vertices. Then how can we obtain a controllable network via synthesizing some leaders and the associated arcs?

The minimal number of leaders required to completely control the overall network was emphasized as an index in some literature [10, 12], which is an integer greater than or equal to 1, denoted by $N_D$. It is supposed that a complex network possesses relatively higher controllability if with lower $N_D$.

As a matter of fact, the $N_D$ of arbitrary real-world networks with continuous-valued arc weights is usually 1.

For the complex network type (2), $N_D$ can be computed according to the following lemma.

*Lemma 4 [3]:* The least number of leaders required for the controllability of dynamical network (2) is the maximum geometric multiplicity of eigenvalues of the matrix $A_{ff}$ representing the follower subgraph.

*Proposition 2*: For any follower subgraph of complex network (2), with the corresponding adjacency matrix $A_{ff} \in R^{N_f \times N_f}$, the probability of $N_D = 1$ is 100%.

*Proof:* According to Lemma 2, for any arbitrary $A_{ff} \in R^{N_f \times N_f}$, the probability of occurrence of the event that maximum algebraic multiplicity of $A_{ff}$ being 1 is 100%. Explicitly, the probability of the event that maximum geometric multiplicity of $A_{ff}$ being 1 is also 100%, because the geometric multiplicity is always less than or equal to the algebraic one. By considering Lemma 4, it can be known that the probability of $N_D = 1$ is 100%. □

If the matrix representation of the complex network is Laplacian type, then it is always true that $N_D = 1$.

*Lemma 5 [3-4]:* For any follower subgraph of complex network (5), with the corresponding Laplacian matrix $L_{ff} \in R^{N_f \times N_f}$, $N_D = 1$.

Since we have known that usually a single leader is sufficient for controllability, let us take a closer look at the cases with one leader, from somewhat different perspective.

*Definition 4:* (Maximum Controllability Index) Consider a dynamical complex network (3) with single leader, where $A_{ff} \in R^{N_f \times N_f}$ is fixed and the remaining blocks are free. The maximum possible dimension of the controllable subspace of the elements in the set $\{(A_{ff}, A_{fl}) \big| A_{fl} \in R^{N_f}\}$ is defined as the *Maximum Controllability Index* of the fixed follower subgraph and is denoted by $\gamma(A_{ff})$.

*Remark 4:* Intuitively, the maximum controllability index can be regarded as the maximum possible quantity of the follower vertices being likely to be under control if with only a single leader.

*Lemma 6 [3]*: The maximum controllability index $\gamma(A_{ff})$ equals the degree of the minimum polynomial of matrix $A_{ff}$.

*Proposition 3:* Almost every $A_{ff} \in R^{N_f \times N_f}$ has the maximum controllability index $\gamma(A_{ff}) = N_f$.

*Proof:* According to Lemma 2, almost any real-valued square matrix $M \in R^{n \times n}$ has maximum algebraic multiplicity of eigenvalues being equal to 1, i.e. for almost any $A_{ff} \in R^{N_f \times N_f}$, the algebraic multiplicities of all the $N_f$ eigenvalues $\lambda_1(A_{ff}), \lambda_2(A_{ff}), \cdots, \lambda_{N_{ff}}(A_{ff})$ are 1. Based on Lemma 3, it can be implied that for almost any $A_{ff} \in R^{N_f \times N_f}$, the degree of minimal polynomial is $N_f$ and then the proof can be concluded with considering Lemma 6.□

*Theorem 2:* For a given $A_{ff} \in R^{N_f \times N_f}$, almost any $A_{fl} \in R^{N_f}$ is effective to render the dimension of controllable subspace of complex network (1) equal to $\gamma(A)$.

*Proof:* For convenience of expression, if the dimension of the controllable subspace of matrix pencil $(A_{ff}, A_{fl})$ is $\gamma(A)$, such a $A_{fl}$ is said to be feasible, otherwise this $A_{fl}$ is infeasible. Suppose that the given matrix $A_{ff} \in R^{N_f \times N_f}$ with Jordan canonical form $J = PA_{ff}P^{-1}$ has $\mu$ distinct eigenvalues $\lambda_1, \lambda_2, ..., \lambda_\mu$. Let $A_{fl} = P^{-1}b$, where $b \in R^{N_f}$ possesses at least $\mu$ non-zero entries, each with index corresponding to the last row of the Jordan block that has the maximum dimension among the Jordan blocks of $\lambda_i$ $(i = 1, 2, ..., \mu)$. Let $q_i^T$ $(i = 1, 2, \cdots, N_f)$ denote the rows in $P^{-1}$ that may be complex valued. If $q_i^T A_{fl} = 0$ for any $i \in \{i_1, i_2, ..., i_\mu\}$, then $A_{fl}$ is infeasible. $q_i^T A_{fl} = 0$ implies both $\mathrm{Re}(q_i^T) A_{fl} = 0$ and $\mathrm{Im}(q_i^T) A_{fl} = 0$, where $\mathrm{Re}(\bullet)$ denotes the real part and $\mathrm{Im}(\bullet)$ the imaginary part. Therefore, the domain in $R^{N_f}$ of infeasible $A_{fl}$ is $\mathrm{N}(\mathrm{Re}(q_i^T)) \cap \mathrm{N}(\mathrm{Im}(q_i^T))$ ( $i = i_1, i_2, ..., i_\mu$ ), where $\mathrm{N}(\bullet)$ denotes the null space. For each $i \in \{i_1, i_2, ..., i_\mu\}$, it is an intersection of two $(N_f - 1)$ -dimensional hyperplanes and must be a subset of measure zero in $R^{N_f}$. Evidently, the complete infeasible domain of $A_{fl}$ is

$$\bigcup_{i \in \{i_1, i_2, ..., i_\mu\}} (\mathrm{N}(\mathrm{Re}(q_i^T)) \cap \mathrm{N}(\mathrm{Im}(q_i^T)))$$

with any vector in the other area of $R^{N_f}$ being feasible. Since this domain is the union of a finite number of subsets of reduced dimension, it is still a null set with measure zero. Thus, almost any vector in the space $R^{N_f}$ of leader-follower arc weights is feasible as $A_{fl}$. □

In fact, almost any general dynamical linear system is controllable, i.e. the probability for any given matrix pencil (*M*, *B*) to be Kalman controllable is 100%, if both *M* and *B* consist of arbitrary parameters; whilst the uncontrollable cases are just the rare exceptions. Here, a simple mathematical proof is presented.

*Proposition 3:* Almost all matrix pencils (*M, B*) with single input variable and arbitrary values of parameters are Kalman controllable.

*Proof:* Suppose that $M \in R^{n \times n}$ and $B \in R^{n \times 1}$. There are $n(n+1)$ free parameters within matrices *M* and *B*. According to Lemma 1, (*M*, *B*) is uncontrollable if and only if the following equation holds

$$\left\| \begin{bmatrix} B & MB & \cdots & M^{n-1}B \end{bmatrix} \right\| = 0$$

Evidently, the elements of matrices $M$ and $B$ satisfying this equation form a manifold of dimension less than $n(n+1)$. Thus, the Lebesgue measure of the set of uncontrollable cases is 0 with regard to the overall parameter space.□

Proposition 3 can be further illustrated by a simple example.

*Example 1:* Consider the network (1) with two followers, which are vertices 1, 2; and one leader, vertex 3. In this case, $A_{ff} \in R^{2\times 2}$ and $A_{fl} \in R^{2\times 1}$, and there are six free parameters: $a_{11}$, $a_{12}$, $a_{21}$, $a_{22}$, $a_{13}$, and $a_{23}$. All the possible values of the six parameters form the space $R^6$. The associated network is uncontrollable if and only if

$$\left|\begin{bmatrix} A_{ff} & A_{ff}A_{fl} \end{bmatrix}\right| = 0$$

Substituting $a_{11}$, $a_{12}$, $a_{21}$, $a_{22}$, $a_{13}$, and $a_{23}$ into the above equation yields

$$\left|\begin{bmatrix} a_{13} & a_{11}a_{13} + a_{12}a_{23} \\ a_{23} & a_{21}a_{13} + a_{22}a_{23} \end{bmatrix}\right| = 0$$

and

$$a_{21}a_{13}^2 + a_{22}a_{13}a_{23} - a_{11}a_{13}a_{23} - a_{12}a_{23}^2 = 0$$

Apparently, the set

$$\left\{\begin{bmatrix} a_{11} & a_{12} & a_{21} & a_{22} & a_{13} & a_{23} \end{bmatrix}^T \middle| a_{21}a_{13}^2 + a_{22}a_{13}a_{23} - a_{11}a_{13}a_{23} - a_{12}a_{23}^2 = 0\right\}$$

is a manifold with dimension less than 6.

The pivotal implication behind the above discussions can be illustrated by Fig. 1 [18].

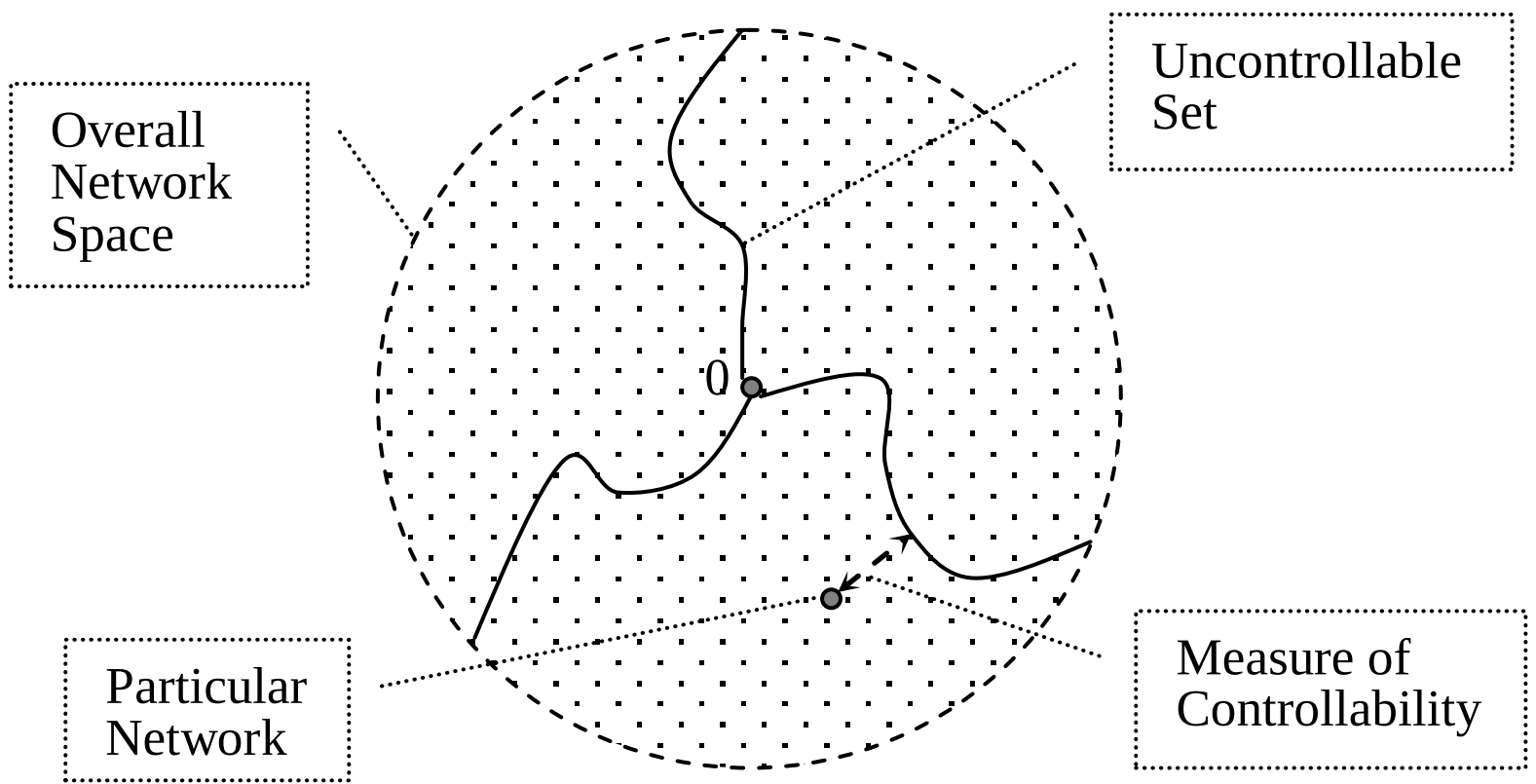

Fig. 1. Network space of $R^{N\times N}$.
Solid curves represent the set of networks that cannot be exactly controlled by only one input signal.

In Fig. 1, any network is represented by a point in the space. Almost all the networks with only one leader vertex are controllable. If a particular network happens to locate in the uncontrollable set, then any infinitesimal perturbation could let it leave this set and become controllable henceforth.

In this regard, a better way to evaluate the intensity of controllability should be to measure how close a given network is to being uncontrollable. Such a route can be called computational controllability [18]. As far as our knowledge is concerned, there are still many problems around computational controllability for general complex networks waiting for solutions.

## 4. Empirical Analysis

In this section, numerical experiments are conducted to validate the almost controllability. Here, The primary manner of experiments is to generate thousands of complex networks with similar settings and observe the percentage of uncontrollable cases appeared. The experiments are performed for three well-known types of complex networks: ER (Erdös and Rényi) network, WS (Watts and Strogatz) small-world network, and BA (Barabási and Albert) scale-free network. All the dynamical networks have a single leader, with the basic arc weight 1, and noise $\varepsilon / k$, where $\varepsilon$ is a random number evenly distributed on [-0.5, 0.5] and $k > 0$ is a constant which determines the magnitude of noise. $k$ is called *noise coefficient*. The magnitude of noise is inversely proportional to the value of $k$.

It is worth mentioning that throughout this section, there are some common settings: the connectivity probability of ER network is $p = 0.4$; for WS small-world network, the switching probability is $p = 0.5$, and $K = 2$; for BA scale-free network, the number of added vertices is $t = 8$, and the number of new edges per vertex is $m = 3$.

Fig. 2 illustrates the relationship between noise coefficient and percentage of uncontrollable cases for three types of complex networks under the dynamical model (1). The network comprises 11 followers and 1 leader for each trial. The noises are unstructured, being independent of the particular structure of any network topology.

It can be seen that the percentage of uncontrollable cases keeps low, whilst it has evident negative correlation with the overall magnitude of noise since it increases with the noise coefficient $k$.

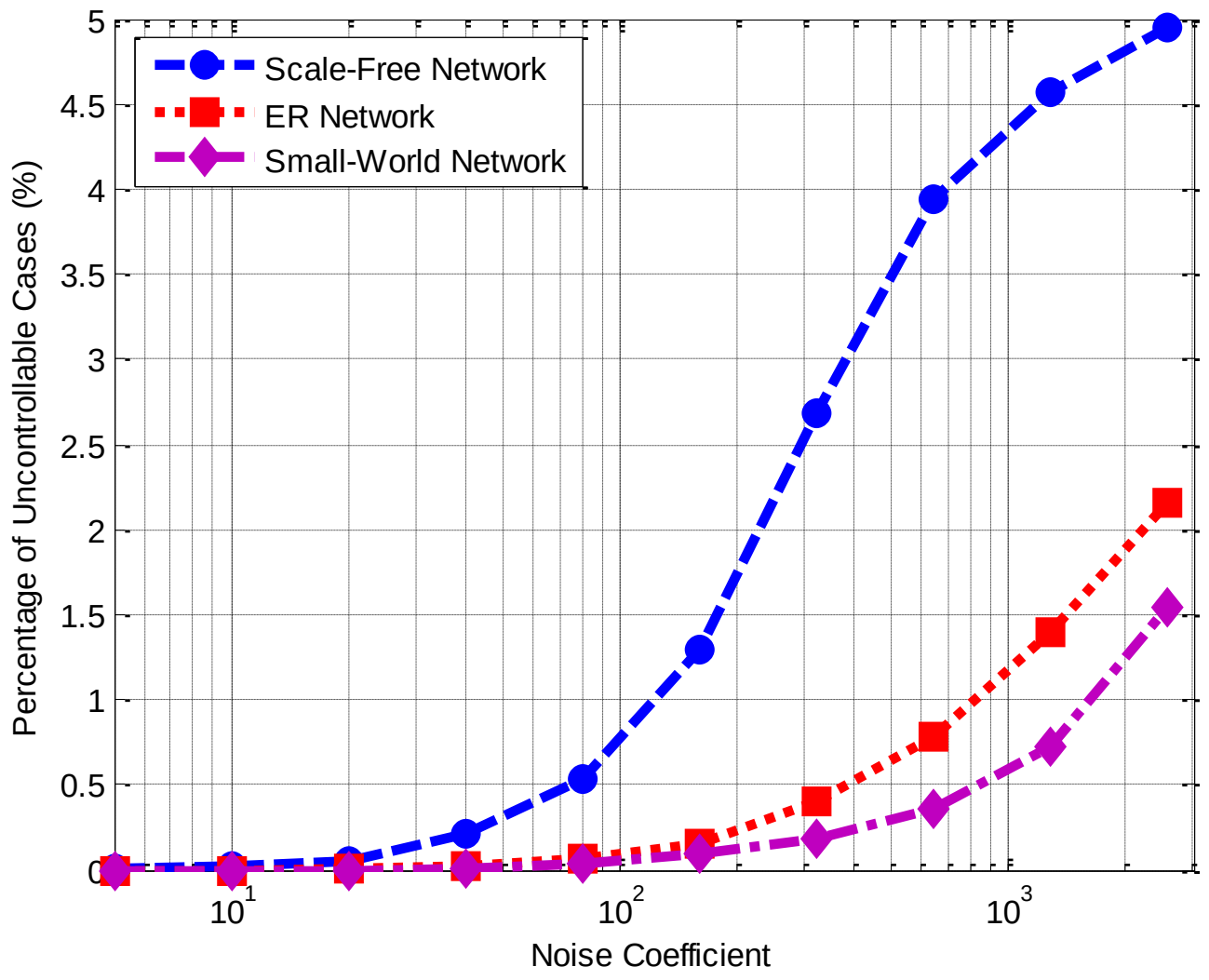

Fig. 2. Trend of percentage of uncontrollable cases for three types of networks represented by adjacency matrix, with 11 followers and unstructured noises.

Fig. 3 illustrates a comparison of the cases with and without structured noise for ER networks described by the dynamical model (1). The noise is deemed structured if the perturbations only dwell on the weighted values of the existing definite edges. One can see that the probability of uncontrollability is higher with structured noises. This is partially because the structural uncontrollability being possibly to occur cannot be broken by structured noises.

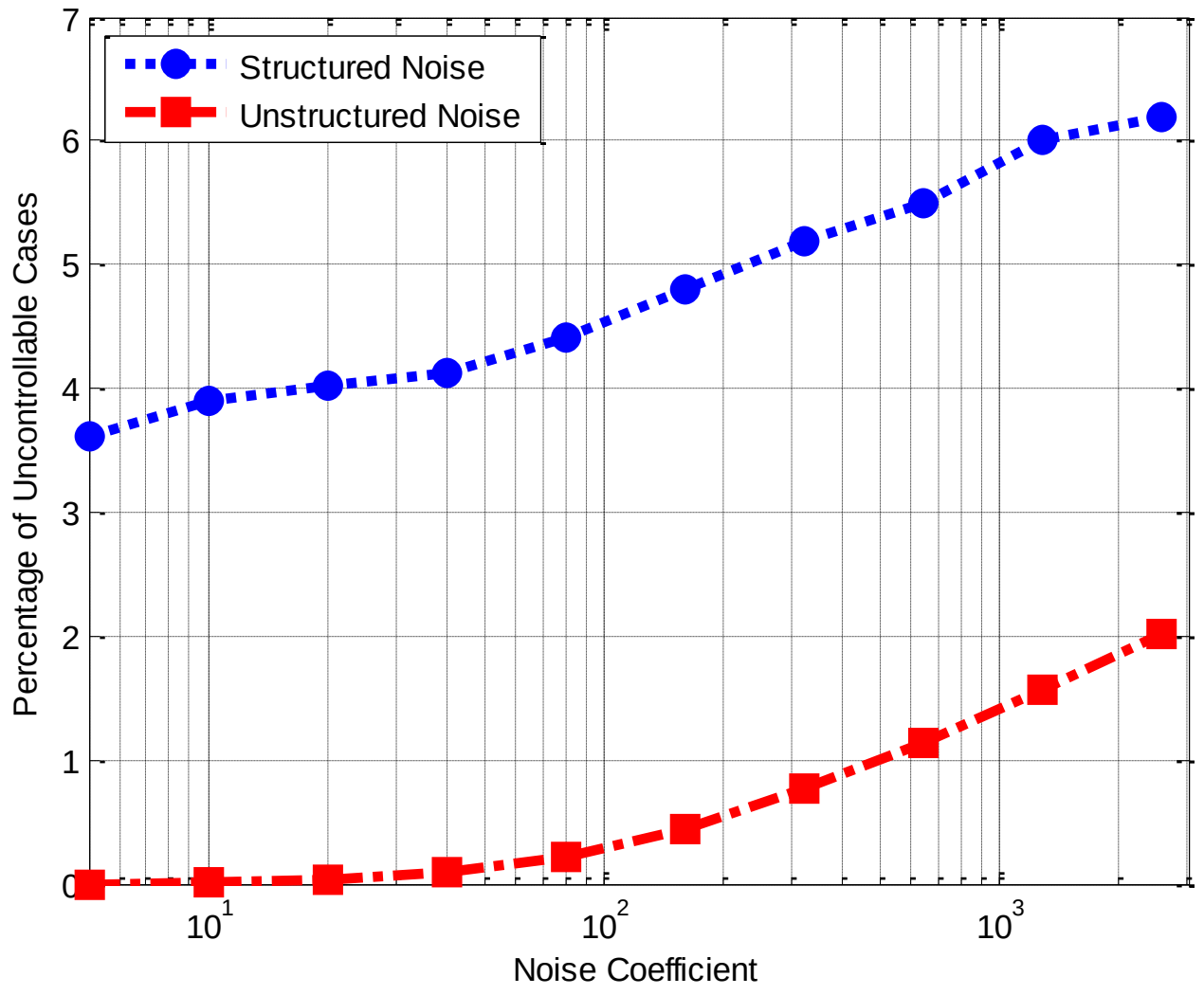

Fig. 3. Trend of percentage of uncontrollable cases for ER networks represented by adjacency matrix, with 12 followers and structured/unstructured noises.

Fig. 4 illustrates the relationship between noise coefficient and percentage of uncontrollable cases for three types of complex networks under the dynamical model (4). The network comprises 10 followers and 1 leader for each trial. The noises are unstructured, being independent of the particular structure of any network topology.

One main difference between Figs 2 & 4 is in the algebraic representation of the complex networks. The networks are represented by adjacency matrix for Fig. 2, whereas the networks are represented by Laplacian matrix for Fig. 4.

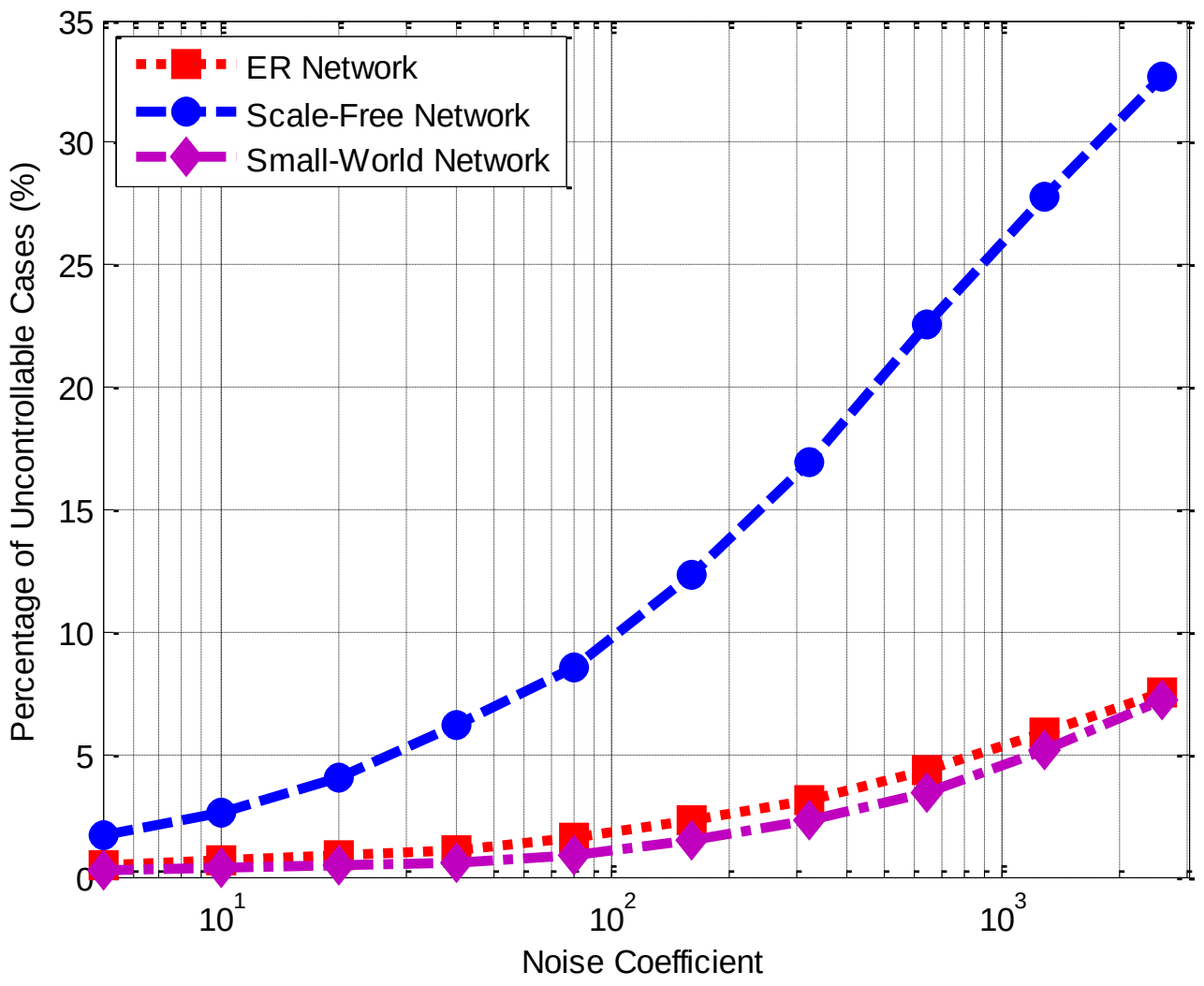

Fig. 4. Trend of percentage of uncontrollable cases for three types of networks represented by Laplacian matrix, with 10 followers and unstructured noises.

It is evident that the probability of uncontrollability is generally higher if the network models are represented by Laplacian matrix, especially for scale-free networks. Furthermore, a specific comparison is illustrated by Fig. 5, which is between small-world networks represented by adjacency and Laplacian matrix, respectively.

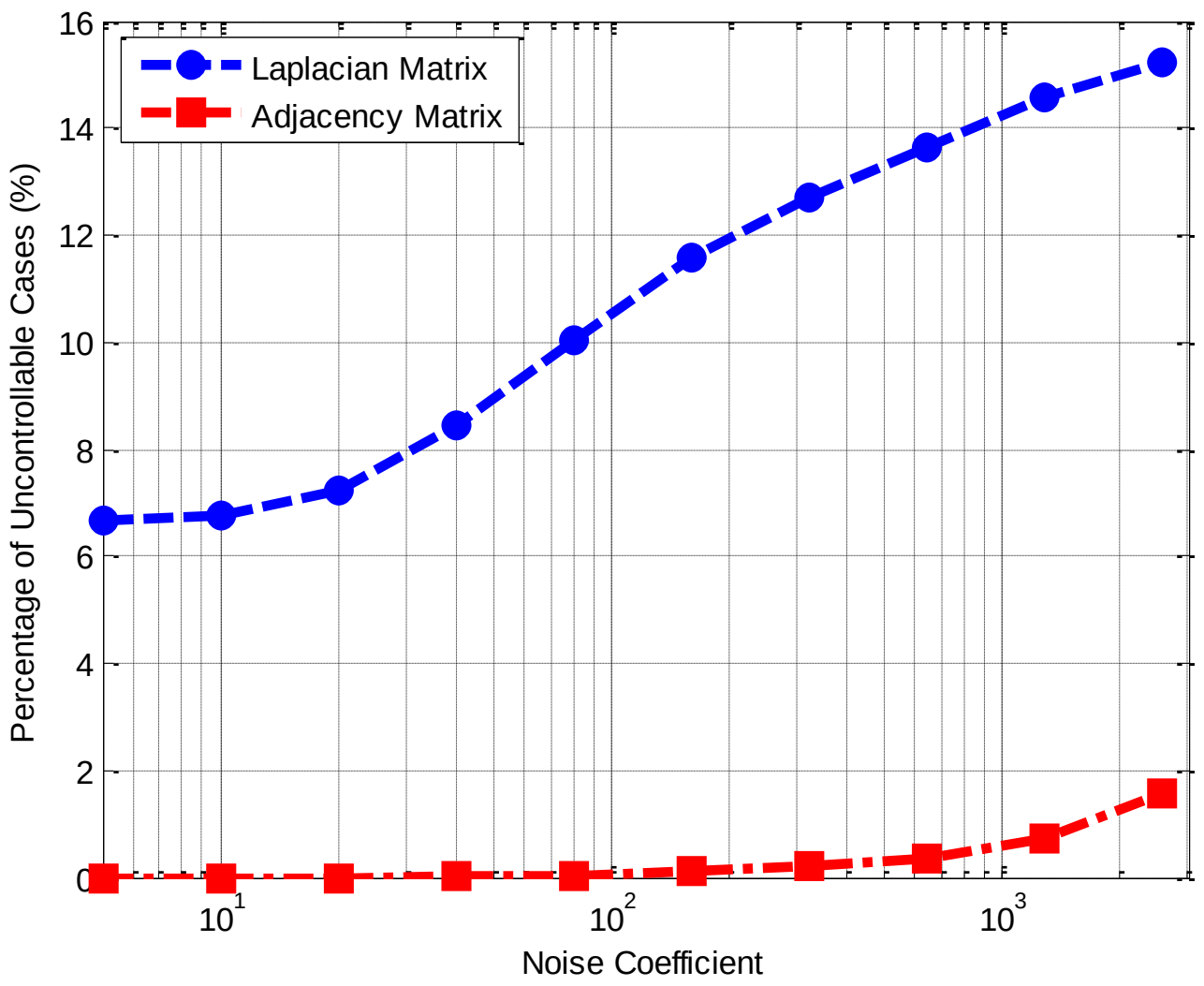

Fig. 5. Trend of percentage of uncontrollable cases for WS small-world networks represented by adjacency/Laplacian matrix, with 11 followers and unstructured noises.

*Remark 5:* It is worth mentioning that the experimental results in this section are numerical, yielded by MATLAB under a default precision setting. Any judgement of a reduced rank is only based on a calculated result that the absolute value of determinant is less than a prescribed small threshold, but not necessarily identical to zero. The results are strongly dependent on the setting of such a threshold. Therefore, one should be cautious to compare networks with different orders by employing the index here, i.e. probability of uncontrollability.

## 5. Conclusion

During the current decade, the controllability problem of dynamical complex networks has received extensive attention, which on one hand has commonness with the controllability of conventional distributed large-scale systems concerned in systems science, whereas on the other hand it also holds some new characteristics. This paper elaborates a notion that weighted real-world networks are almost controllable in the sense of Kalman controllability, if bearing modeling deviations. The concept of almost controllability here may include three main facets: for the networks with noises on the weighted arcs, 1) the probability for the requisite number of independent input signals being 1 is 100%; 2) the probability for networks with single input signal being controllable is 100%; 3) an uncontrollable network would become controllable by any infinitesimal perturbation. Intuitively speaking, uncontrollability is not robust against deviations in models, whereas controllability is. Two types of dynamical models are addressed to show almost controllability, which are represented by adjacency and Laplacian matrices respectively. Both theoretical and empirical analyses are presented in detail to elucidate the notion. As the further studies along this route, presently at least two noticeable directions exist: 1) deeper explorations on computational controllability for real-world complex networks could be conducted; 2) certain phenomena observed in experiments could possibly be explained in analytic manner, e.g. why scale-free networks appear relatively prone to be uncontrollable.

## Acknowledgments

This work is supported by National Natural Science Foundation (NNSF) of China

(Grants 61374054 & 61263002), by the Zhejiang Open Foundation of the Most Important Subjects, by Fundamental Research Funds for the Central Universities (Grant 31920160003), and by Program for Young Talents of State Ethnic Affairs Commission (SEAC) of China (Grant 2013-3-21).

## Conflict of Interest

The authors declare that there is no conflict of interest regarding the publication of this paper.

## Appendix

The subsequent theoretical grounds are essential for proving Lemma 2, which are quoted from linear algebra.

*Definition 5 [22]:* (Resultant) Suppose that $f(x) = a_0x^m + a_1x^{m-1} + \cdots + a_m$ ( $a_0 \neq 0$ ) and $g(x) = b_0x^n + b_1x^{n-1} + \cdots + b_n$ ( $b_0 \neq 0$ ), with their roots being $\alpha_1, \alpha_2, \cdots, \alpha_m$ and $\beta_1, \beta_2, \cdots, \beta_n$ respectively. The resultant of the two polynomials is defined as follows:

$$R(f,g) = a_0^n b_0^m \prod_{i=1}^{m} \prod_{j=1}^{n} (\alpha_i - \beta_j) \tag{8}$$

With the following formula [22], one can directly compute the resultant of two polynomials by their coefficients:

$$R(f,g) = \Delta(m,n) = \begin{vmatrix} a_0 & a_1 & \cdots & a_m & & & & \\ & a_0 & a_1 & \cdots & a_m & & & \\ & & & & \ddots & & & \\ & & & & a_0 & a_1 & \cdots & a_m \\ b_0 & b_1 & \cdots & b_n & & & & \\ & b_0 & b_1 & \cdots & b_n & & & \\ & & & & \ddots & & & \\ & & & & b_0 & b_1 & \cdots & b_n \end{vmatrix}$$

*Lemma 6 [22]:* An irreducible polynomial $p(x)$ is a multiple factor of polynomial $f(x)$ if and only if $p(x)$ is a common factor of $f(x)$ and $f'(x)$, where $f'(x)$ is the derivative of $f(x)$.

*Lemma 7 [22]:* There exists a common factor of polynomials $f(x)$ and $g(x)$

if and only if the resultant $R(f,g)=0$.

*Proof of Lemma 2:* Consider any arbitrary matrix $M=[m_{ij}]\in R^{n\times n}$. Suppose that its characteristic polynomial is

$$f(x)=|xI-M|=x^n+c_1x^{n-1}+\cdots+c_n \tag{9}$$

with the coefficient vector $\begin{bmatrix}c_1 & c_2 & \cdots & c_n\end{bmatrix}^T\in R^n$. The derivative of $f(x)$ is

$$f'(x)=nx^{n-1}+c_1(n-1)x^{n-2}+\cdots+c_{n-1}$$

According to Lemmas 6 & 7, there exists a multiple factor of $f(x)$ if and only if $R(f,f')=0$, i.e. the coefficients of $f(x)$ must satisfy the following equation:

$$\begin{vmatrix}
1 & c_1 & \cdots & c_n & & & & \\
 & 1 & c_1 & \cdots & c_n & & & \\
 & & & & \ddots & & & \\
 & & & & 1 & c_1 & \cdots & c_n \\
n & c_1(n-1) & \cdots & c_{n-1} & & & & \\
 & n & c_1(n-1) & \cdots & c_{n-1} & & & \\
 & & & & \ddots & & & \\
 & & & & n & c_1(n-1) & \cdots & c_{n-1}
\end{vmatrix}=0 \tag{10}$$

Evidently, the set of variables $c_1,\cdots,c_n$ satisfying the above algebraic equation forms a manifold of dimension $n-1$.

Equation (10) determines the relationship between the coefficient vector $\begin{bmatrix}c_1 & c_2 & \cdots & c_n\end{bmatrix}^T$ and the values of elements in matrix *M*, which are *n* scalar functions:

$$\begin{cases}
c_1=\varphi_1(m_{11} \ \cdots \ m_{1n} \ m_{21} \ \cdots \ m_{2n} \ \cdots \ m_{n1} \ \cdots \ m_{nn}) \\
c_2=\varphi_2(m_{11} \ \cdots \ m_{1n} \ m_{21} \ \cdots \ m_{2n} \ \cdots \ m_{n1} \ \cdots \ m_{nn}) \\
\quad \vdots \\
c_n=\varphi_n(m_{11} \ \cdots \ m_{1n} \ m_{21} \ \cdots \ m_{2n} \ \cdots \ m_{n1} \ \cdots \ m_{nn})
\end{cases} \tag{11}$$

Substituting $c_1,\cdots,c_n$ in (10) by (11) will lead to an equation with the form:

$$\Gamma(m_{11} \ \cdots \ m_{1n} \ m_{21} \ \cdots \ m_{2n} \ \cdots \ m_{n1} \ \cdots \ m_{nn})=0 \tag{12}$$

where $\Gamma: R\times R\times\cdots\times R\to R$ is a nonlinear function constraining the values of elements by some addition and multiplication operations. It is evident that the set of matrices that satisfy (12) constitute a nonlinear manifold of reduced dimension in $R^{n\times n}$ and therefore its Lebesgue measure is zero. All the matrices in $R^{n\times n}$ except this set possess no multiple eigenvalues. □

## References

[1] H. G. Tanner, "On the controllability of nearest neighbor interconnections", *Proc. 43rd IEEE Conf. Decision and Control*, vol. 3, pp. 2467-2472, 2004.

[2] A. Rahmani, M. Ji, and M. Mesbahi *et al.*, "Controllability of multi-agent systems from a graph-theoretic perspective", *SIAM J. Control Optim.*, vol. 48, no.1, pp. 162-186, 2009.

[3] N. Cai, J.-X. Xi, and Y.-S. Zhong *et al.*, "Controllability improvement for linear time-invariant dynamical multi-agent systems", *Int. J. Innov. Comput. I.*, vol. 8, no. 5a, pp. 3315-3328, 2012.

[4] N. Cai and M. J. Khan, "On generalized controllability canonical form with multiple input variables", *Int. J. Control Automat. Syst.*, vol. 15, pp. 169-177, 2017.

[5] N. Cai, J. Zhou, and L. Guo, "Almost exact controllability of dynamic complex networks", *Proc. Chin. Control Conf.*, pp. 254-257, 2016.

[6] B. Liu, T.-G. Chu, and L. Wang *et al.*, "Controllability of a leader-follower dynamic network with switching topology", *IEEE Trans. Automat. Control*, vol. 53, no.4, pp. 1009-1013, 2008.

[7] Z.-J. Ji and H.-S. Yu, "A new perspective to graphical characterization of multi-agent controllability", *IEEE Trans. Cybernet.*, DOI: 10.1109/TCYB.2016.2549034.

[8] Z.-J. Ji, H. Lin, and H.-S. Yu, "Protocols design and uncontrollable topologies construction for multi-agent networks", *IEEE Trans. Automat. Control*, vol. 60, no. 3, pp. 781-786, 2015.

[9] Y.-Q. Guan, Z.-J. Ji, L. Zhang, and L. Wang, "Controllability of multi-agent systems under directed topology", *Int. J. Robust Nonlin. Control*, DOI: 10.1002/rnc.3798.

[10] Y.-Y. Liu, J. J. Slotine, and A. L. Barabási, "Controllability of complex networks", *Nature*, vol. 473, pp. 167-173, 2011.

[11] G. Yan, J. Ren, and Y.-C. Lai *et al.*, "Controlling complex networks: How much energy is needed?", *Phys. Rev. Lett.*, vol. 108, 218703, 2012.

[12] Z.-Z. Yuan, C. Zhao, and Z.-R. Di *et al.*, "Exact controllability of complex networks", *Nat. Commn.*, vol. 4, 2013.

[13] J. Sun and A. E. Motter, "Controllability transition and nonlocality in network control", *Phys. Rev. Lett.*, vol. 110, 208701, 2013.

[14] F. Pasqualetti, S. Zampieri, and F. Bullo, "Controllability metrics, limitations and algorithms for complex networks", *IEEE Trans. Control Netw. Syst.*, vol. 1, pp. 40-52, 2014.

[15] T. Jia, Y.-Y. Liu, and E. Csóka *et al.*, "Emergence of bimodality in controlling complex networks", *Nat. Commun.*, vol. 4, 2002, 2013.

[16] Y.-Y. Liu, J. J. Slotine, and A. L. Barabási, "Control centrality and hierarchical structure in complex networks", *PLoS ONE*, vol. 7, no. 9, e44459, 2012.

[17] J. Ruths and D. Ruths, "Control profiles of complex networks", *Science*, vol. 21, pp. 1373-1376, 2014.

[18] N. Cai, "On quantitatively measuring controllability of complex networks", *Physica A*, vol. 474, pp. 282-292, 2017.

[19] T. Kailath, *Linear Systems*, Englewood Cliffs: Prentice-Hall, 1980.

[20] J.-X. Xi, M. He, and H. Liu *et al.*, "Admissible output consensualization control for singular multi-agent systems with time delays", *J. Franklin Inst.*, vol. 353, pp. 4074-4090, 2016.

[21] N. Cai and M. J. Khan, "Almost decouplability of any directed weighted network topology", *Physica A*, vol. 436, pp. 637-645, 2015.

[22] B. Feng, *Polynomials and Irrational Numbers*, Harbin: Harbin Institute of Technology Press, 2008. (In Chinese)